\documentclass[preprint,onecolumn,nofootinbib]{revtex4}
\usepackage{graphicx}
\usepackage{amsmath}
\usepackage{amsfonts}
\usepackage{amssymb}
\usepackage[normalem]{ulem}
\usepackage{color,xcolor}
\usepackage{CJK}
\usepackage{subfigure}
\usepackage{amsthm}
\usepackage{mathrsfs}
\usepackage{url}
\usepackage{array}
\usepackage{dcolumn}
\usepackage{float}
\usepackage{appendix}
\usepackage{multirow}
\usepackage{booktabs}
\usepackage[colorlinks=true,linkcolor=blue,urlcolor=blue,filecolor=black,citecolor=red,pdfstartview=FitV,pdftitle={},pdfsubject={},pdfkeywords={},pdfpagemode=UseNone,bookmarksopen=true,hypertexnames=false]{hyperref}

\begin{document}
\title{Echoes and quasinormal modes for static loop quantum black bounces}

\author{Huajie Gong$^{1}$}
\thanks{huajiegong@hunnu.edu.cn}
\author{Guoyang Fu$^{2}$}
\author{Shulan Li $^{3}$}
\author{Jian-Pin Wu$^{2}$}
\thanks{jianpinwu@yzu.edu.cn, corresponding author}
\author{Qin Tan$^{1}$}
\author{Qiyuan Pan$^{1}$}
\thanks{panqiyuan@hunnu.edu.cn, corresponding author}

\affiliation{
	$^{1}$ \mbox{Department of Physics, Institute of Interdisciplinary Studies, Key Laboratory of} \mbox{Low Dimensional Quantum Structures and Quantum Control of Ministry of Education,} \mbox{Synergetic Innovation Center for Quantum Effects and Applications, and Hunan Research} \mbox{Center of the Basic Discipline for Quantum Effects and Quantum Technologies,} \mbox{Hunan Normal University, Changsha 410081, China}\\
	$^{2}$ \mbox{Center for Gravitation and Cosmology, College of Physical Science and Technology,} \mbox{Yangzhou University, Yangzhou 225009, China}\\
	$^{3}$ \mbox{Department of Physics, Shanghai University, Shanghai 200444, China}\\
}

\begin{abstract}
We investigate scalar perturbations of the static loop quantum black bounce (LQBB) spacetime with multipole index $l=1$, focusing on time-domain signals and fundamental quasinormal frequencies (QNFs). The LQBB model provides a unified description of regular black holes (RBHs) and traversable wormholes, governed by the quantum parameter $\alpha$ and the bounce parameter $r_b$. Using the finite difference method, we find no echoes for the displayed RBH configurations with a single-barrier effective potential, whereas clear echoes are produced by the potential well structure in selected traversable wormhole configurations. The QNFs obtained from the Prony method and the direct integration method are in good agreement. In the RBH case, increasing $r_b$ or $\alpha$ leads to a slower decay. In the wormhole case, the QNFs depend non-monotonically on the model parameters, and the emergence of echoes is closely tied to the effective potential profile. These results show that the LQBB spacetime provides a useful framework for studying wave dynamics in RBHs and traversable wormholes, and for clarifying how horizon and throat structures affect ringdown and echoes.
\end{abstract}

\maketitle
\tableofcontents

\section{Introduction}
Recent observational breakthroughs in gravitational wave astronomy by the LIGO-Virgo-KAGRA collaborations and in black hole (BH) imaging by the Event Horizon Telescope (EHT) have opened unprecedented windows for testing fundamental theories of gravity in the strong field regime \cite{LIGOScientific:2016aoc, LIGOScientific:2017vwq, EventHorizonTelescope:2022wkp, EventHorizonTelescope:2019dse}. These high-precision observations not only confirm the predictions of general relativity (GR) in numerous scenarios but also motivate rigorous explorations beyond GR to resolve long-standing theoretical puzzles, most notably the spacetime singularity problem inherent in classical BH solutions \cite{PhysRevLett.14.57,hawking1970singularities}. As a promising non-perturbative quantum gravity framework, loop quantum gravity (LQG) has been successfully applied to cosmological models and provides a natural pathway to regularize spacetime singularities by introducing quantum corrections at the Planck scale \cite{Rov,Thiemann:2001gmi,Ashtekar:2004eh,Han:2005km,Yang:2022aec,Han:2024ydv,Zhang:2024ney}.

Motivated by LQG quantization techniques for spherically symmetric spacetimes, a quantum-corrected BH solution has been constructed \cite{Kelly:2020uwj}. Later, in the quantum Oppenheimer-Snyder (qOS) model \cite{Lewandowski:2022zce}, the effective exterior metric, which is matched to the collapsing dust ball described by loop quantum cosmology, reproduces the earlier LQG-corrected BH solution and shares its characteristic minimal radius $r_m$ that resolves the central singularity. We refer to this quantum-corrected BH solution as the qOS BH henceforth. This model has attracted attention as a concrete background for studying the observational and dynamical imprints of loop quantum corrections, including BH shadows, images, quasinormal modes (QNMs), periodic orbits and so on \cite{Yang:2022btw,Zhang:2023okw,Gong:2023ghh,Yang:2024lmj}. However, this solution still allows the radial coordinate to extend to $r\to0$, leaving the spacetime singularity unresolved. To address this limitation, a natural strategy is to combine the LQG-corrected geometry with a global regularization mechanism. The Simpson-Visser black bounce construction provides precisely such a mechanism. By replacing $r$ with $\sqrt{r^2+r_b^2}$, it converts the Schwarzschild center into a regular bounce and gives a one-parameter family that interpolates between the Schwarzschild BH, regular black holes (RBHs), one-way wormholes, and traversable wormholes \cite{Simpson:2018tsi}. This interpolation changes the horizon and throat structure, as well as the photon sphere, innermost stable circular orbit (ISCO), and Regge-Wheeler potential of the spacetime. Consequently, black bounce geometries have become useful backgrounds for studying the transition between RBHs and wormholes, together with the associated QNMs and echo signals \cite{Churilova:2019cyt,Yang:2021cvh,Wu:2022eiv,Yang:2022ryf}.

Building on these two developments, the loop quantum black bounce (LQBB) spacetime proposed in Ref.~\cite{Muniz:2024wiv} combines the qOS quantum-corrected geometry with the Simpson-Visser black bounce regularization. In this construction, the parameter associated with loop quantum corrections and the bounce parameter jointly determine whether the geometry describes an RBH or a traversable wormhole. The model therefore retains the quantum-corrected character of the qOS BH while replacing the central singularity by a globally regular bounce, and it offers a single framework in which horizon geometries and throat geometries can be compared. This feature is directly relevant for the present work, because the appearance of echoes and the behavior of QNMs are controlled by the effective potential, which is sensitive to whether the spacetime contains an event horizon, a throat, or a potential well. Existing studies of shadows and accretion disk images have further shown that LQBB parameters can leave observable imprints \cite{He:2025hbu}, while the perturbative ringdown and echo properties of this spacetime remain a natural subject for further investigation.

QNMs are the characteristic decaying oscillations of compact objects under perturbations. Their complex frequencies are entirely determined by the intrinsic properties of spacetime and act as a vital probe for exploring strong field gravity \cite{LIGOScientific:2021sio,Berti:2015itd,Berti:2018vdi,Cardoso:2019rvt,Fu:2022cul,Fu:2023drp,Gong:2023ghh,Xia:2023zlf,Guo:2023nkd,Zhang:2024nny,Song:2024kkx,Yang:2024prm,Tan:2024qij,Dong:2024ams,Deng:2025hfn}. Notably, gravitational wave echoes may appear in the late time tail of the ringdown. They originate from the multiple reflections of perturbation waves in the potential well of horizonless exotic compact objects (ECOs) such as wormholes, gravastars, and boson stars \cite{Cardoso:2016oxy,Hui:2023ibl,Zhang:2023mzb,Yang:2022ryf,Liu:2021aqh,Abedi:2016hgu,Conklin:2017lwb}. As a key observable, echoes can effectively distinguish classical BHs from horizonless ECOs and provide a new approach to testing the fundamental nature of gravity \cite{Cardoso:2017cqb,Mark:2017dnq,Konoplya:2018yrp,LongoMicchi:2019wsh,Churilova:2019cyt,Bronnikov:2019sbx,DuttaRoy:2019hij,Chowdhury:2020rfj,Chowdhury:2022zqg,Zhang:2025ygb}.

Against this background, we analyze the scalar field perturbations of the static LQBB spacetime, focusing on the time-domain evolution of perturbation waves and the fundamental quasinormal frequencies (QNFs). We employ the finite difference method (FDM) \cite{Abdalla:2010nq,Zhu:2014sya,Lin:2022rtx} to simulate the perturbation evolution and extract echo signals, and subsequently adopt the Prony method \cite{Berti:2007dg,Konoplya:2011qq} and the direct integration method (DIM) \cite{Ferrari:2007rc,Pani:2012bp,Pierini:2023btw} to compute the QNFs systematically. Our main results show that clear echo signals emerge for selected configurations in the traversable wormhole regime, whereas no echoes arise for the RBH configurations within the parameter ranges studied. This feature makes the LQBB spacetime a promising platform for probing quantum gravity effects and exploring the nature of compact objects.

The paper is organized as follows. Section~\ref{LQBB-model} presents the LQBB model and the scalar field perturbation equations. Section~\ref{Veff-LQBB} investigates the effective potential. Section~\ref{Echo-LQBB} discusses scalar wave echoes from the time-domain method. Section~\ref{QNMs-LQBB} computes and discusses the QNFs. Section~\ref{sec-conclusion} summarizes the conclusions.

\section{The LQBB model and its scalar field perturbations}\label{LQBB-model}
Recent work in Ref.~\cite{Muniz:2024wiv} proposed a static black bounce geometry incorporating LQG corrections, also referred to as the LQBB spacetime. The metric is given by
\begin{equation}\label{eq2}
	ds^2=-f(r) dt^2 + f(r)^{-1}dr^2+(r^2+r_b^2) d\Omega^2,
\end{equation}
with
\begin{equation}\label{fr}
	f(r)= 1-\frac{2M}{\sqrt{r^2+r_b^2}}+\frac{\alpha^2 M^2}{(r^2 + r_b^2)^2},
\end{equation}
where $d\Omega^2=d\theta^2+\sin^{2}\theta d\phi^2$ is the line element on a unit sphere, and $M$ denotes the mass. In principle, we treat $\alpha$ and $r_b$ as free parameters.

This model represents a class of regular spacetime solutions aimed at resolving the singularity problem in GR. Its theoretical construction proceeds in two key steps:

1. By applying the quantization techniques of LQG to a spherically symmetric spacetime, one obtains a static, spherically symmetric BH solution incorporating loop quantum corrections \cite{Kelly:2020uwj,Lewandowski:2022zce}. Its metric is $ds^2 =-(1 - 2M/r + \alpha^2 M^2/r^4) dt^2 + (1 - 2M/r + \alpha^2 M^2/r^4)^{-1} dr^2 + r^2 d\Omega^2$, where the parameter $\alpha$ encodes the quantum effects from LQG. Although this solution is valid only above a minimum radius $ r_m = (\alpha^2 M/2)^{1/3} $, the metric itself can be analytically extended to the $ r \rightarrow 0 $ region, which remains characterized by a singularity. Consequently, it fails to eliminate the spacetime singularity entirely.

2. To globally remove the singularity, the model incorporates the Simpson-Visser prescription \cite{Simpson:2018tsi} by implementing the substitution $ r \rightarrow \sqrt{r^2 + r_{b}^2} $, with a nonzero regularization parameter $r_{b}$ (also known as the bounce parameter), thereby extending the radial coordinate to the entire real domain. The LQBB metric is subsequently derived when the regularization parameter is chosen as $r_b = r_m$.

The defining feature of this model lies in its globally nonsingular spacetime structure. It can smoothly interpolate between an RBH and a traversable wormhole depending on the values of the mass $M$, the LQG parameter $\alpha$, and the bounce parameter $r_b$. Notably, the resulting solutions can be interpreted as spacetime structures supported by sources described by nonlinear electrodynamics (NED) and a scalar field within the framework of GR, which provides a new avenue for constructing semiclassical models that are both nonsingular and physically more realistic~\cite{Muniz:2024wiv}.

For convenience, we define $\rho = \sqrt{r^2 + r_b^2}$ and take $r_b>0$ \footnote{The spacetime geometry is symmetric under a sign change of $r_b$ because it enters the metric as $r_b^2$. Consequently, the solutions for negative $r_b$ are physically equivalent to those for positive $r_b$, and it is sufficient to consider $r_b > 0$.}. The metric function $f(r)$ can then be expressed as
\begin{equation}
g(\rho) = 1 - \frac{2M}{\rho} + \frac{\alpha^2 M^2}{\rho^4}.
\end{equation}
Setting $g(\rho)=0$, one obtains the inner horizon $\rho_{-}$ and the outer horizon $\rho_{+}$.

Depending on the quantum parameter $\alpha$ and the bounce parameter $r_b$, this spacetime can be classified into the following two categories:
\begin{itemize}
	\item \textbf{RBH:} this case occurs when $\alpha \leq \sqrt{27/16}M$ and the bounce parameter satisfies $r_b < \rho_{+}$;
	\item \textbf{Traversable wormhole:} this case exists in two parameter regimes:
	\begin{enumerate}
		\item $\alpha \leq \sqrt{27/16}M$ with the bounce parameter satisfying $r_b > \rho_{\pm}$;
		\item $\alpha > \sqrt{27/16}M$ for any positive bounce parameter $r_b > 0$.
	\end{enumerate}
\end{itemize}

\begin{figure}[htbp]
	\centering
	\includegraphics[scale = 1]{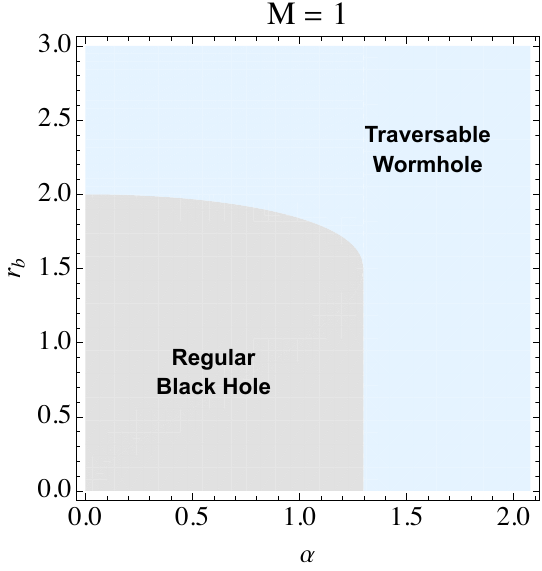}
	\caption{Parameter space of the LQBB spacetime for $M=1$.}
	\label{Fig0}
\end{figure}

As shown in Fig.~\ref{Fig0}, the light-gray region corresponds to RBHs, which occur for $\alpha\leq\sqrt{27/16}M$ and $r_b<\rho_+$. The light-blue region represents traversable wormholes, including both the regime $\alpha\leq\sqrt{27/16}M$ with $r_b>\rho_\pm$ and the regime $\alpha>\sqrt{27/16}M$ with any $r_b>0$. Without loss of generality, we set $M=1$ throughout this paper.

To investigate the dynamical properties of this spacetime, we consider the propagation of a massless scalar field. The covariant Klein-Gordon equation for a massless scalar field is given by
\begin{equation}
	 \frac{1}{\sqrt {-g}}\partial_{\mu}(\sqrt {-g}g^{\mu\nu}\partial_{\nu}\Psi )=0 . \label{KGeq} 
\end{equation}
We decompose $\Psi(t,r,{\theta},{\phi})$ into spherical harmonics as
\begin{equation}
	\Psi(t,r,{\theta},{\phi})=\frac{1}{\sqrt{r ^{2}+r_{b}^{2}}}\Phi(r,t)Y^{m}_{l}({\theta},{\phi}),\label{dp}
\end{equation}
where $Y^{m}_{l}({\theta},{\phi})$ is the spherical harmonic function of degree $l$ (multipole index) and order $m$ (azimuthal index), satisfying the angular eigenvalue equation
\begin{equation}
	[\frac{1}{\sin{\theta}}\frac{\partial}{\partial{\theta}}(\sin{\theta}\frac{\partial}{\partial{\theta}})+\frac{1}{\sin^{2}{\theta}}   \frac{\partial^{2}}{\partial{\phi}^{2}}]Y^{m}_{l}({\theta},{\phi})=-l(l+1)Y^{m}_{l}({\theta},{\phi}). \label{AEE}       
\end{equation}
Substituting Eq.~\eqref{dp} into Eq.~\eqref{KGeq} yields the Schr\"odinger-like wave equation
\begin{equation}
	[\frac{\partial^{2}}{\partial{t}^{2}}-\frac{\partial^{2}}{\partial r^{2}_{*}}+V(r)]\Phi(r,t)=0, \label{Seq}  
\end{equation}
with $V(r)= f(r)	\left(\left(r^2+r_{b}^2\right) \left(r f'(r)+l^2+l\right)+r_{b}^2 f(r)\right)/\left(r^2+r_{b}^2\right)^2$ and the tortoise coordinate $r_{*}$ defined by $r_{*}=\int \frac{dr}{f(r)}$.

\section{The effective potential for the LQBB model}\label{Veff-LQBB}
It is well known that the shape of the effective potential plays a crucial role in the behavior of both echoes and QNMs \cite{Fu:2023drp,Gong:2023ghh,Zhang:2025ygb}. In this section, we analyze the effective potential for scalar field perturbations in the LQBB spacetime in both the BH and traversable wormhole cases. Throughout this work, we fix $l = 1$.

\begin{figure}[htbp]
	\centering
	\includegraphics[scale = 0.8]{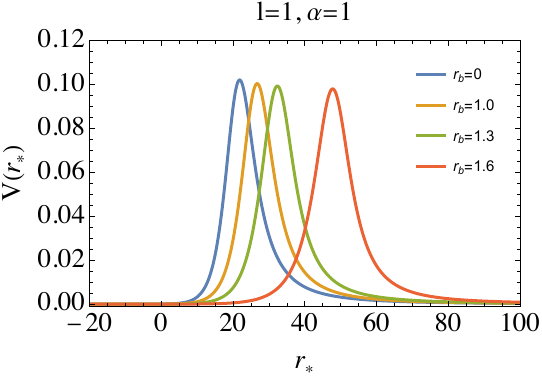}
	\includegraphics[scale = 0.8]{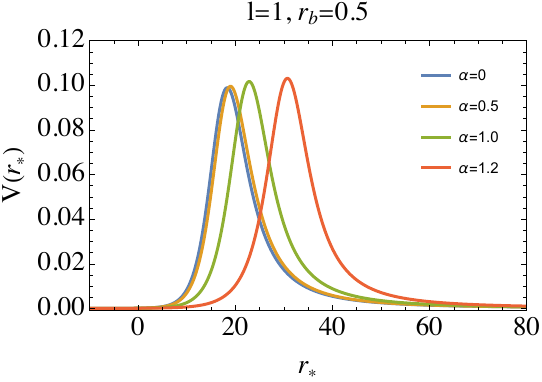}
	\caption{The effective potential $V(r_*)$ for perturbations of the scalar field on the RBH spacetime is shown for different values of $r_{b}$ with $\alpha=1$ in the left panel and different values of $\alpha$ with $r_{b}=0.5$ in the right panel, respectively.}
	\label{FigveffBH}
\end{figure}
For the BH case, Fig.~\ref{FigveffBH} illustrates the effective potential of scalar field perturbations for different values of $\alpha$ and $r_b$. The effective potential remains positive everywhere, implying that this RBH spacetime is stable against scalar field perturbations. In the left panel of Fig.~\ref{FigveffBH}, we fix $\alpha = 1$ and vary $r_b$. The peak of the effective potential gradually decreases as $r_b$ increases. In the right panel, we fix $r_b = 0.5$ and vary $\alpha$; the peak value then increases with $\alpha$. These opposite influences of $\alpha$ and $r_b$ on the effective potential suggest that they encode different physics. Owing to this contrast, we expect distinct signatures to emerge later in the echo and QNM analyses. Moreover, the effective potentials of the RBH remain finite in all cases and exhibit a single peak only, consistent with the description in Ref.~\cite{Wu:2022eiv}.

\begin{figure}[htbp]
	\centering
	\includegraphics[scale = 0.8]{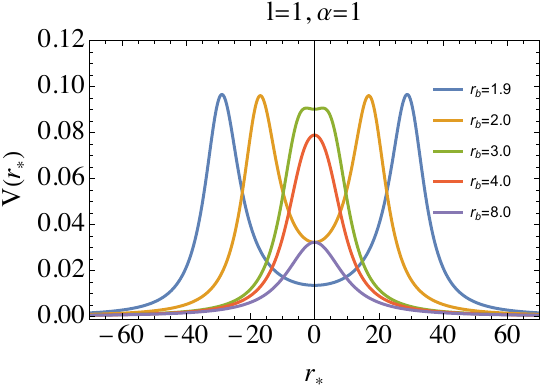}
	\includegraphics[scale = 0.8]{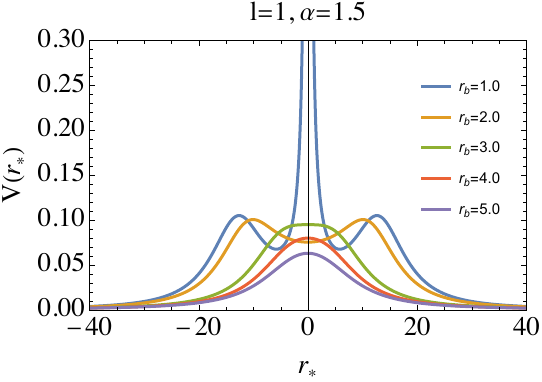}
	\caption{The effective potential $V(r_*)$ for perturbations of the scalar field on the traversable wormhole spacetime is shown for different values of $r_{b}$ with $\alpha=1$ in the left panel and $\alpha=1.5$ in the right panel, respectively.}
	\label{FigveffWH1}
\end{figure}

\begin{figure}[htbp]
	\centering
	\includegraphics[scale = 0.75]{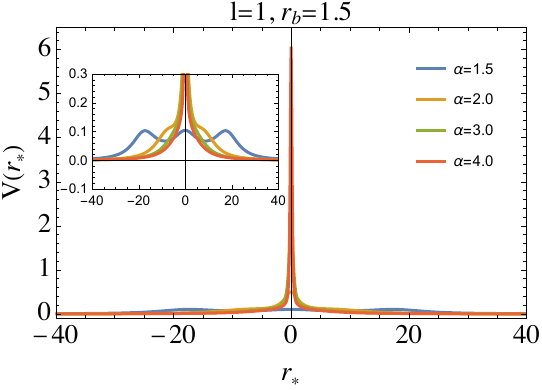}
	\includegraphics[scale = 0.8]{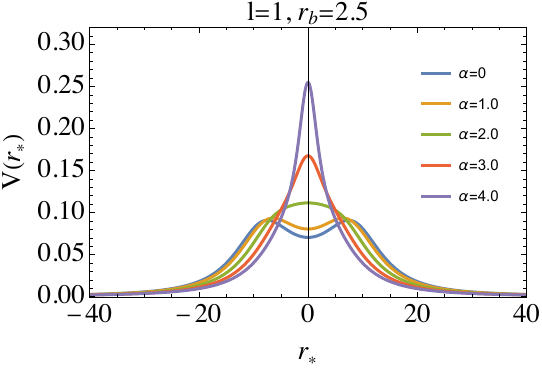}
	\caption{The effective potential $V(r_*)$ for perturbations of the scalar field on the traversable wormhole spacetime is shown for different values of $\alpha$ with $r_{b}=1.5$ in the left panel and $r_{b}=2.5$ in the right panel, respectively.}
	\label{FigveffWH2}
\end{figure}

For traversable wormholes, we also investigate the behavior of the effective potential under scalar field perturbations as a function of $r_b$ and $\alpha$, as shown in Figs.~\ref{FigveffWH1} and \ref{FigveffWH2}. In Fig.~\ref{FigveffWH1}, we fix $\alpha = 1$ (left panel) and $\alpha = 1.5$ (right panel) to analyze the influence of $r_b$. The choice of these two values of $\alpha$ is motivated by the different ranges of $r_b$ required for a wormhole spacetime: in the left panel, $r_b > 1.83929$ is necessary for $\alpha = 1$, whereas in the right panel, $r_b > 0$ suffices for $\alpha = 1.5$. This choice ensures the generality of our analysis. In both panels, as $r_b$ increases, the effective potential transitions from a multi-peak structure to a single-peak profile. Such multi-peak features have also been reported in the literature \cite{Yang:2021cvh}.

The potential well structure observed in Fig.~\ref{FigveffWH1} is a key ingredient for the occurrence of echoes. We therefore expect to detect echo signals under such parameter configurations. In particular, for the right panel of Fig.~\ref{FigveffWH1}, as $ r_b \rightarrow 0 $, the effective potential diverges at $ r_* $. This behavior can be understood from the metric expression \eqref{fr}: when $ r = 0 $ (i.e., the throat of the wormhole) and $ r_b \rightarrow 0 $, the metric itself tends to diverge, consequently leading to the divergence of the corresponding effective potential.

Next, we turn to Fig.~\ref{FigveffWH2} to examine the influence of the quantum parameter $\alpha$ on the effective potential of the traversable wormhole. For the same reasons discussed previously, we consider two representative values, $r_{b} = 1.5$ and $r_{b} = 2.5$. As $\alpha$ increases, the effective potential gradually transitions from a multi-peak structure to a single-peak structure, consistent with the earlier analysis. However, in the single-peak regime, the height of the potential peak increases with $\alpha$, in contrast to the behavior shown in Fig.~\ref{FigveffWH1}, where the peak value decreases as $r_b$ increases. For the case with a potential well in the left panel of Fig.~\ref{FigveffWH2} (e.g., when $\alpha = 1.5$), this structure is favorable for echoes, consistently with the corresponding time-domain waveform; whereas for the right panel of Fig.~\ref{FigveffWH2}, the potential well is relatively shallow, and any resulting echo is expected to be weak and may not be clearly distinguishable in the time-domain waveform \cite{Tan:2024qij}.

\section{Echoes for the LQBB model}\label{Echo-LQBB}
In this section, we focus on the time-domain evolution of the LQBB model under scalar field perturbations. For convenience, we present the results with the same parameters as those in the previous section. Detailed calculations employed for this analysis can be found in Appendix~\ref{TDI-method}.

\begin{figure}[htbp]
	\centering
	\includegraphics[scale = 0.8]{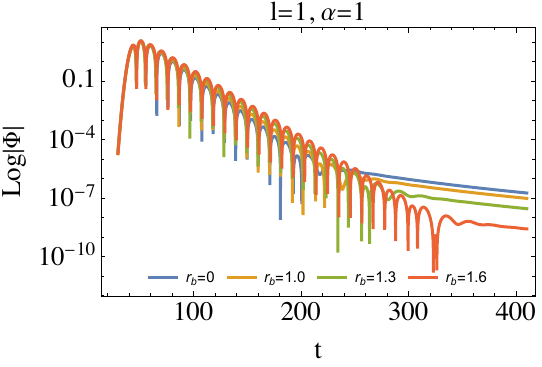}
	\includegraphics[scale = 0.8]{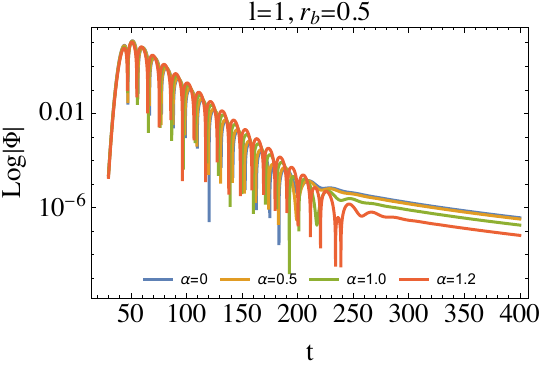}
	\caption{The time evolution of RBH for perturbations of the scalar field is shown for different values of $r_{b}$ with $\alpha=1$ in the left panel and different values of $\alpha$ with $r_{b}=0.5$ in the right panel, respectively.}
	\label{Figecho1}
\end{figure}
Fig.~\ref{Figecho1} presents the time-domain evolution of the RBH in the LQBB model under scalar field perturbations. No echo is observed, as expected from the structure shown in Fig.~\ref{FigveffBH}, where the effective potential exhibits a single-peak profile for the displayed cases. For the waveforms displayed here, the RBH ringdown decays more slowly than the qOS BH ($r_b=0$) and black-bounce ($\alpha=0$) reference curves. Furthermore, increasing the bounce parameter $r_b$ with fixed $\alpha$, or increasing $\alpha$ with fixed $r_b$, leads to a more slowly decaying ringdown.

\begin{figure}[htbp]
	\centering
	\includegraphics[scale = 0.8]{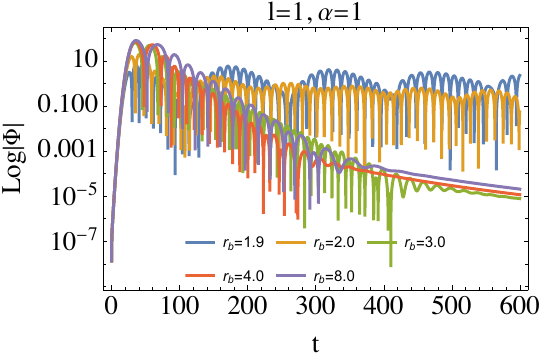}
	\includegraphics[scale = 0.8]{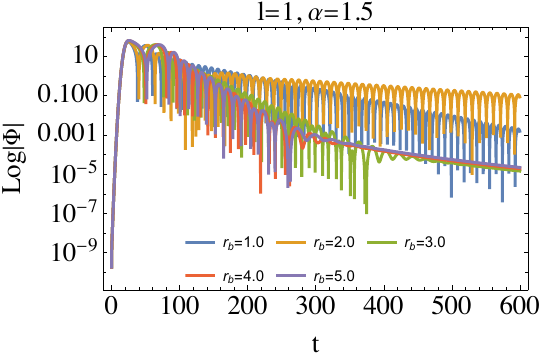}
	\caption{The time evolution of the traversable wormhole for perturbations of the scalar field is shown for different values of $r_{b}$ with $\alpha=1$ in the left panel and $\alpha=1.5$ in the right panel, respectively.}
	\label{Figecho2}
\end{figure}

\begin{figure}[htbp]
	\centering
	\includegraphics[scale = 0.8]{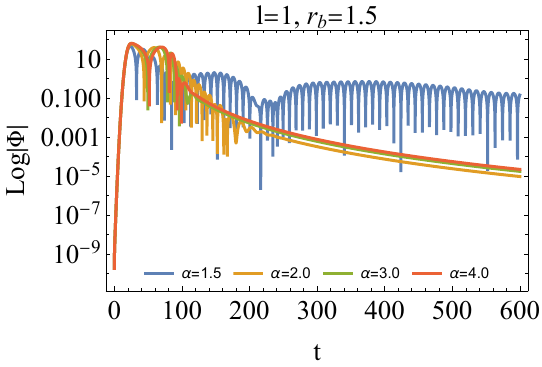}
	\includegraphics[scale = 0.8]{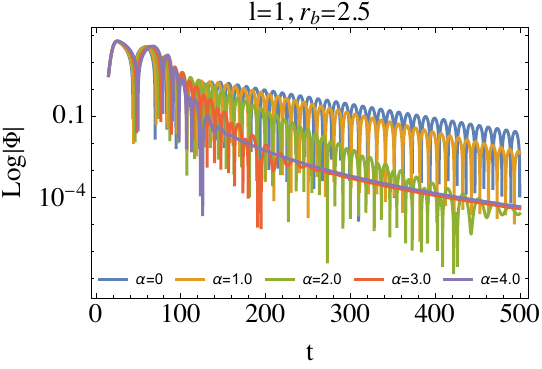}
	\caption{The time evolution of the traversable wormhole for perturbations of the scalar field is shown for different values of $\alpha$ with $r_{b}=1.5$ in the left panel and $r_{b}=2.5$ in the right panel, respectively.}
	\label{Figecho3}
\end{figure}

We next present the time-domain evolution results for scalar perturbations in the wormhole scenario within the LQBB model in Figs.~\ref{Figecho2} and \ref{Figecho3}. The main conclusions are as follows:
\begin{enumerate}
	\item Echoes are observed for selected configurations in the traversable wormhole regime of this model. They are clearly visible in the left panels of Figs.~\ref{Figecho2} and \ref{Figecho3}, consistent with the analysis in Sec.~\ref{Veff-LQBB}.
	
	\item In the left panel of Fig.~\ref{Figecho2}, as $r_b$ increases with fixed $\alpha=1$, the time interval between echo signals becomes shorter, and the echo signals gradually weaken and eventually disappear. This is because the width of the potential well in Fig.~\ref{FigveffWH1} becomes smaller, finally forming a single-peaked potential barrier. This conclusion is consistent with that in Ref.~\cite{Wu:2022eiv}. Similarly, in the left panel of Fig.~\ref{Figecho3}, increasing $\alpha$ with fixed $r_b=1.5$ leads to the same trend, showing that the echoes become progressively weaker and are no longer clearly distinguishable as $\alpha$ increases.
	
\end{enumerate}

\section{QNFs for the LQBB model}\label{QNMs-LQBB}
In this section, we present the numerical results for the QNFs of the model. We adopt two methods to compute the QNFs, namely the Prony method and the DIM\footnote{The detailed computational procedures can be found in Appendix~\ref{DI-method}.}. When solving for QNMs, the boundary conditions for BHs and wormholes exhibit both similarities and notable differences.

Hereafter, we adopt the convention of harmonic time dependence $e^{-i\omega t}$, such that the sign of the imaginary part of the QNF follows the standard convention, with $\mathrm{Im}(\omega)<0$ corresponding to decaying modes. For BHs, the QNM boundary conditions require the wave function to be a purely incoming wave at the event horizon and a purely outgoing wave at spatial infinity:
\begin{equation}
\text{$\Phi_{RBH}$} \propto
\begin{cases}
\text{$e^{-i\omega r_*}$}, & r_*\to-\infty,\\[2pt]
\text{$e^{+i\omega r_*}$}, & r_*\to+\infty,
\end{cases}
\end{equation}
where $r_{*}$ is the tortoise coordinate.

For wormholes, despite the differences in spacetime structure and geometry compared with BHs, the QNM boundary conditions remain similar. The two endpoints of the wormhole correspond to two different asymptotic regions in space, and the boundary conditions therefore require the wave function to be purely outgoing in both asymptotic regions:
\begin{equation}
\text{$\Phi_{WH}$} \propto
\begin{cases}
\text{$e^{-i\omega r_*}$}, & r_*\to-\infty,\\[2pt]
\text{$e^{+i\omega r_*}$}, & r_*\to+\infty.
\end{cases}
\end{equation}

Unlike BHs, wormholes do not have an event horizon. Instead, the throat connects the two asymptotic regions. Since the effective potential is symmetric under $r_*\to -r_*$, the scalar perturbations can be decomposed into even and odd parity sectors. The even sector satisfies $\left.d\Phi_{WH}/dr_*\right|_{r_*=0}=0$ at the throat, while the odd sector satisfies $\left.\Phi_{WH}\right|_{r_*=0}=0$.

In summary, despite the differences in the spacetime structures of BHs and wormholes, their QNM boundary conditions are similar in terms of the outgoing wave requirement, and thus the same numerical methods can be used to solve for the QNMs in both cases. In particular, for wormholes, the outgoing condition is imposed at both asymptotic infinities, while the throat condition is determined by the parity sector. Since the effective potential is symmetric about the throat, the perturbations can be decomposed into independent even- and odd-parity sectors. In this work, we focus on the even-parity sector and accordingly adopt a static even-parity initial wave packet, which is symmetric about the throat and has a vanishing initial time derivative.

\begin{table}[htbp]
	\caption{Fundamental QNFs of the RBH case of the LQBB spacetime for $l=1$ and $M=1$.}
	\label{tb-A}
	\centering
	\scalebox{1}{
		\begin{tabular}{|c|l|c|c|}
			\hline
			\multicolumn{2}{|c|}{} & \textbf{Prony} & \textbf{DIM} \\ 
			\hline
			\multirow{4}{*}{$\alpha=1.0$} 
			& $r_{b}=0$ & 0.299179 - 0.0922678i & 0.299179 - 0.0922571i \\
			\cline{2-4}
			& $r_{b}=1.0$ & 0.298410 - 0.0866014i & 0.298426 - 0.0866256i \\ 
			\cline{2-4}
			& $r_{b}=1.3$ & 0.297322 - 0.0808969i & 0.297642 - 0.0825820i \\ 
			\cline{2-4}
			& $r_{b}=1.6$ & 0.296359 - 0.0772640i & 0.295469 - 0.0770260i \\ 
			\hline
			\multirow{4}{*}{$r_{b}=0.5$} 
			& $\alpha=0$ & 0.292883 - 0.0962666i & 0.292874 - 0.0962529i \\
			\cline{2-4}
			& $\alpha=0.5$ & 0.294394 - 0.0951983i & 0.294350 - 0.0951800i \\
			\cline{2-4}
			& $\alpha=1.0$ & 0.299016 - 0.0908763i & 0.299008 - 0.0908710i \\
			\cline{2-4}
			& $\alpha=1.2$ & 0.301318 - 0.0872701i & 0.300900 - 0.0875341i \\ 
			\hline
		\end{tabular}
	}
\end{table}

\begin{table}[htbp]
	\caption{Fundamental QNFs of the wormhole case of the LQBB spacetime for $l=1$, $M=1$, fixed $\alpha$, and varying $r_b$.}
	\label{tb-B}
	\centering
	\scalebox{1}{
		\begin{tabular}{|c|l|c|c|}
			\hline
			\multicolumn{2}{|c|}{} & \textbf{Prony} & \textbf{DIM} \\ 
			\hline
			\multirow{5}{*}{$\alpha=1.0$} 
			& $r_{b}=1.9$ & echo & 0.314188 - 0.0069937i, echo \\
			\cline{2-4}
			& $r_{b}=2.0$ & echo & 0.329620 - 0.0154331i, echo \\
			\cline{2-4}
			& $r_{b}=3.0$ & 0.313454 - 0.0346959i & 0.313450 - 0.0346947i \\
			\cline{2-4}
			& $r_{b}=4.0$ & 0.278335 - 0.0483499i & 0.280467 - 0.0499645i \\ 
			\cline{2-4}
			& $r_{b}=8.0$ & 0.168621 - 0.0431197i & 0.170187 - 0.0450417i \\ 
			\hline
			\multirow{5}{*}{$\alpha=1.5$}  
			&$r_{b}=1.0$  & 0.334223 - 0.0159742i & 0.333739 - 0.0152344i \\
			\cline{2-4}
			&$r_{b}=2.0$  & 0.301843 - 0.0053689i & 0.301846 - 0.0053677i \\
			\cline{2-4}
			&$r_{b}=3.0$  & 0.317951 - 0.0378888i & 0.317947 - 0.0378877i \\
			\cline{2-4}
			&$r_{b}=4.0$  & 0.281308 - 0.0507759i & 0.281309 - 0.0508076i \\
			\cline{2-4}
			&$r_{b}=5.0$  & 0.244584 - 0.0527503i & 0.244597 - 0.0527683i \\
			\hline
		\end{tabular}
	}
\end{table}

\begin{table}[htbp]
	\caption{Fundamental QNFs of the wormhole branch of the LQBB spacetime for $l=1$, $M=1$, fixed $r_b$, and varying $\alpha$.}
	\label{tb-c}
	\centering
	\scalebox{1}{
		\begin{tabular}{|c|l|c|c|}
			\hline
			\multicolumn{2}{|c|}{} & \textbf{Prony} & \textbf{DIM} \\ 
			\hline
			\multirow{4}{*}{$r_{b}=1.5$}  
			&$\alpha=1.5$  & echo & 0.292600 - 0.0022223i, echo \\
			\cline{2-4}
			&$\alpha=2.0$  & 0.402693 - 0.0849738i & 0.402802 - 0.0814630i \\
			\cline{2-4}
			&$\alpha=3.0$  & 0.373818 - 0.2027247i & 0.379997 - 0.1999999i \\
			\cline{2-4}
			&$\alpha=4.0$  & 0.320995 - 0.2473835i & 0.329301 - 0.2490001i \\
			\hline
			\multirow{5}{*}{$r_{b}=2.5$} 
			& $\alpha=0$ & 0.297448 - 0.0120491i & 0.297446 - 0.0120494i \\
			\cline{2-4}
			& $\alpha=1.0$ & 0.310297 - 0.0166078i & 0.310293 - 0.0166074i \\
			\cline{2-4}
			&$\alpha=2.0$  & 0.342317 - 0.0332326i & 0.342311 - 0.0332330i \\
			\cline{2-4}
			&$\alpha=3.0$  & 0.378411 - 0.0676870i & 0.381744 - 0.0677882i \\
			\cline{2-4}
			&$\alpha=4.0$  & 0.420520 - 0.1270904i & 0.419850 - 0.1268640i \\
			\hline
		\end{tabular}
	}
\end{table}

The QNFs corresponding to different quantum parameters $\alpha$ and bounce parameters $r_b$ are listed in Table~\ref{tb-A} for the RBH case and in Tables~\ref{tb-B} and~\ref{tb-c} for the wormhole case. Our data show a consistent trend with the time-domain waveforms depicted in Figs.~\ref{Figecho1}-\ref{Figecho3}. Based on these data, the results are summarized as follows:
\begin{enumerate}
	\item The QNFs from the Prony and DIM methods agree well, with only tiny numerical discrepancies, confirming the reliability and self-consistency of our computations.
	
	\item For the RBH case (see Table~\ref{tb-A}), with fixed quantum parameter $\alpha=1.0$, as the bounce parameter $r_b$ increases, the real part of the QNF decreases slightly, while the absolute value of the imaginary part decreases significantly. This indicates that an increase in $r_b$ reduces the oscillation frequency and slows the decay of the BH perturbation, leading to a longer-lived ringdown. For fixed bounce parameter $r_b=0.5$, as the quantum parameter $\alpha$ increases, the real part of the QNF increases monotonically, while the absolute value of the imaginary part decreases gradually. This implies that a larger $\alpha$ enhances the oscillation and suppresses the energy dissipation during the ringdown stage.
	
	\item Tables~\ref{tb-B} and~\ref{tb-c} show that both $r_b$ and $\alpha$ affect the wormhole QNFs and the appearance of echoes. Echoes are identified at $r_b=1.9$ and $2.0$ for $\alpha=1.0$, and at $r_b=1.5$ for $\alpha=1.5$. At fixed $\alpha$, both the real part and the magnitude of the imaginary part exhibit a non-monotonic dependence on $r_b$ within the sampled data. For fixed bounce parameter $r_b$, the real part still behaves non-monotonically as $\alpha$ increases, while the absolute value of the imaginary part increases monotonically with $\alpha$, meaning that the scalar perturbation of the wormhole decays faster.

	\item A comparison of Tables~\ref{tb-A} and~\ref{tb-c} reveals distinct decay behaviors with respect to $\alpha$ between the RBH and wormhole cases. For the RBH case at fixed $r_b=0.5$ in Table~\ref{tb-A}, the magnitude of the imaginary part decreases as $\alpha$ increases, corresponding to a slower decay. For the wormhole case at fixed $r_b$ in Table~\ref{tb-c}, it generally increases with $\alpha$, corresponding to a faster decay. This contrast suggests the influence of the quantum correction parameter on the mode decay may depend on the underlying spacetime structure, in particular, whether the spacetime contains a throat or an event horizon.
	
\end{enumerate}

\section{Conclusion and discussion}\label{sec-conclusion}
In this work, we have studied scalar field perturbations of the static LQBB spacetime for the multipole index $l=1$ and examined their time-domain waveforms and fundamental QNFs. The LQBB model provides a simple but rich framework that includes both RBH and traversable wormhole configurations, governed by the loop quantum parameter $\alpha$ and the bounce parameter $r_b$.

For the RBH case considered here, the effective potential exhibits a positive single-barrier structure, suggesting scalar-field modes stability. In this regime, the time-domain signal does not show any echo. The waveform is dominated by the ordinary ringdown phase, and its decay becomes slower when either $r_b$ or $\alpha$ increases. This behavior is also reflected in the QNF: the real part changes only mildly, while the magnitude of the imaginary part decreases, indicating a slower decay.

For the traversable wormhole case, the situation is different. The effective potential may develop a potential well between two barriers, and such a structure is favorable for repeated wave reflections. For selected wormhole configurations with a sufficiently pronounced potential well, clear echo signals appear in the time-domain evolution. Within the sampled parameter range, the echoes become weaker and eventually are no longer clearly distinguishable as either $r_b$ or $\alpha$ increases and the potential well becomes shallower. The QNFs of the wormhole also show a nontrivial dependence on the parameters, and the appearance of echoes is closely related to the detailed shape of the effective potential.

We have also computed the fundamental QNFs using two independent numerical methods, namely the Prony method and the DIM. The two methods yield consistent results, supporting the robustness of our numerical analysis. In the wormhole case, the QNFs vary non-monotonically with the model parameters. This contrasts with the RBH case and reflects the different wave dynamics caused by the distinct spacetime structures.

Overall, our results show that the LQBB spacetime can produce qualitatively different scalar field ringdown signatures in the RBH and traversable wormhole regimes. The presence or absence of echoes, together with the behavior of the QNFs, provides a useful test field probe for distinguishing these two geometries and offers guidance for future studies of electromagnetic and gravitational perturbations relevant to observational tests. It would be interesting to extend this study to other types of perturbations, such as electromagnetic and gravitational perturbations, as well as to investigate the rotating version of the LQBB spacetime. These extensions may help build a more comprehensive understanding of the observational signatures of compact objects with quantum gravity corrections.

\appendix

\section{The time-domain integration method}\label{TDI-method}
The time-domain evolution of perturbations has proven to be an effective approach for extracting the ringdown waveform of compact objects, and the finite difference method (FDM) has been extensively applied in this context~\cite{Abdalla:2010nq,Zhu:2014sya,Lin:2022rtx}. The time and tortoise radial coordinates are discretized on a uniform grid with step sizes $\Delta t$ and $\Delta r_*$, such that $t = i \Delta t$ and $r_* = j \Delta r_*$, where the indices $i$ and $j$ denote the number of time and space grid points, respectively. Under this discretization, the field function $\Phi$ and the effective potential take the form
\begin{equation}
\Phi(t,r_*) = \Phi_{i,j}, \quad V_{eff}(r) = V_j \, .
\end{equation}
Substituting the discrete expressions into the wave equation yields the following finite difference representation:
\begin{equation}
	-\frac{(\Phi_{i+1,j}-2\Phi_{i,j}+\Phi_{i-1,j})}{\Delta t^2}+\frac{(\Phi_{i,j+1}-2\Phi_{i,j}+\Phi_{i,j-1})}{\Delta r_*^2}-V_j \Phi_{i,j}=0,
\end{equation}
which can be rewritten in an iterative form as
\begin{equation}
	\Phi_{i+1,j}=-\Phi_{i-1,j}+\frac{\Delta t^2}{\Delta r_*^2}(\Phi_{i,j+1}+\Phi_{i,j-1})+(2-2\frac{\Delta t^2}{\Delta r_*^2}-\Delta t^2 V_j)\Phi_{i,j}\,.
	\label{iteration}
\end{equation}

Numerical stability of the evolution requires that the Courant--Friedrichs--Lewy (CFL) condition be satisfied:
\begin{equation}
\Delta t / \Delta r_* < 1 \, .
\end{equation}
We remark that the stability and accuracy depend not only on this ratio but also on the individual values of $\Delta t$ and $\Delta r_*$. In the present paper, we set $\Delta t / \Delta r_*=1/2$.

Because the field decay is almost independent of the initial conditions, we assume the initial condition to be a Gaussian wave packet \cite{Moderski:2001tk,Moderski:2001gt,Moderski:2005hf}:
\begin{equation}
\Phi(r_*,t<0)=0,\quad \Phi(r_*,t=0)=\exp\!\left[-\frac{(r_*-a)^2}{2b^2}\right],
\end{equation}
after which Eq.~\eqref{iteration} is iterated to obtain the complete time-domain waveform. It should be noted that, for the wormhole case, we symmetrize the Gaussian initial profile about the throat and set its initial time derivative to zero. The resulting static profile has even parity, satisfies the even-parity throat condition, and therefore excites only even-parity modes.

\section{Direct Integration Method}\label{DI-method}
The direct integration method (DIM) is a classical numerical approach for solving eigenvalue problems in differential equations. It has been widely applied in BH perturbation theory, for instance, in the calculation of QNFs \cite{Ferrari:2007rc,Pani:2012bp}. The essence of this method is to transform the boundary value problem into an initial value problem by integrating from the spatial boundaries and imposing continuity conditions at the matching point, thereby determining the eigenvalues (i.e., the QNFs).

Specifically, we consider the second-order wave equation for the radial function $\Phi$
\begin{equation}\label{eqs_appdendix}
	\Phi''(r)+A_1(r)\Phi'(r)+B_1(r)\Phi(r)=0\,,
\end{equation}
with
\begin{equation}
	A_1(r)=f'(r)/f(r)\,, \ \ B_1(r)=(\omega^2-V(r))/f(r)^2\,.
\end{equation}
Here, $f(r)$ is the spacetime metric function, $\omega=\omega_R+i\omega_I$ is the QNF (with $\omega_R$ as the oscillation frequency and $\omega_I$ as the damping rate), and $V(r)$ is the perturbation's effective potential.

To solve for the QNMs of a BH, we impose ingoing and outgoing wave conditions at the event horizon ($r=r_h$) and spatial infinity ($r=\infty$), respectively:
\begin{eqnarray}
	\label{bcs_hor_appendix}
	\Phi&=& \sum_{i=0}^{N_H}e^{-i \omega r_*} {\Phi}_{i}^{H}(\omega,r)\quad \text{for }  r\rightarrow
	 r_h,\  \\ 
	\label{bcd_inf_appendix}
	\Phi&=& \sum_{i=0}^{N_\infty}e^{+i\omega r_*} \Phi_{i}^{\infty}(\omega,r)\quad \text{for }  r\rightarrow
	 \infty.
\end{eqnarray}
Here, $N_H$ and $N_{\infty}$ denote the series expansion orders of $\Phi_{i}^{H}(\omega,r)$ and $\Phi_{i}^{\infty}(\omega,r)$ at their corresponding boundaries. It should be emphasized that the truncation order $N_{\infty}$ needs to be chosen large enough to suppress numerical spurious effects caused by using the finite radial boundary $r_{\text{inf}}$ to approximate infinity.

For the wormhole, the boundary condition essentially involves connecting two copies of the same geometry via a thin shell and satisfying the continuity condition for perturbation wave functions at the junction (i.e., the throat) \cite{Cardoso:2016oxy}. This fundamentally differs from the absorbing boundary condition of a BH, where only ingoing waves are permitted at the horizon. Since the effective potential is symmetric under $r_*\to -r_*$, the scalar perturbations can be decomposed into even and odd parity sectors. Using this symmetry, one may solve the problem only on one side of the throat. In the present work, we focus on the even-parity sector, for which the throat condition is given by:
	\begin{equation}
		\left.\frac{d\Phi}{dr_*}\right|_{r_*=0}=0 \, .
	\end{equation}
For the odd parity sector, the corresponding throat condition is
	\begin{equation}
		\left.\Phi\right|_{r_*=0}=0 \, .
	\end{equation}
Thus, the full wormhole QNM spectrum contains both even and odd parity sectors. In the present calculation, we impose the even parity condition in the DIM and the odd parity sector can be computed in the same way by replacing the throat condition with $\left.\Phi\right|_{r_*=0}=0$.

\section*{Acknowledgments}

We are very grateful to Xi-Jing Wang, Dan Zhang and Zhong-Wu Xia for helpful discussions and suggestions. This work is supported by the Natural Science Foundation of China (Grants Nos. 12275079, 12035005, 12405055, 12347111 and 12375055), the Key Project of the Department of Education of Hunan Province (Grant No. 25A0084), the innovative research group of Hunan Province (2024JJ1006) and the China Post-doctoral Science Foundation (Grant No. 2025M773339).

\bibliographystyle{style1}
\bibliography{Ref}

@article{LIGOScientific:2016aoc,
	author = "Abbott, B. P. and others",
	collaboration = "LIGO Scientific, Virgo",
	title = "{Observation of Gravitational Waves from a Binary Black Hole Merger}",
	eprint = "1602.03837",
	archivePrefix = "arXiv",
	primaryClass = "gr-qc",
	reportNumber = "LIGO-P150914",
	doi = "10.1103/PhysRevLett.116.061102",
	journal = "Phys. Rev. Lett.",
	volume = "116",
	number = "6",
	pages = "061102",
	year = "2016"
}

@article{LIGOScientific:2017vwq,
	author = "Abbott, B. P. and others",
	collaboration = "LIGO Scientific, Virgo",
	title = "{GW170817: Observation of Gravitational Waves from a Binary Neutron Star Inspiral}",
	eprint = "1710.05832",
	archivePrefix = "arXiv",
	primaryClass = "gr-qc",
	reportNumber = "LIGO-P170817",
	doi = "10.1103/PhysRevLett.119.161101",
	journal = "Phys. Rev. Lett.",
	volume = "119",
	number = "16",
	pages = "161101",
	year = "2017"
}

@article{EventHorizonTelescope:2022wkp, 
	author = "Akiyama, Kazunori and others",
	collaboration = "Event Horizon Telescope",
	title = "{First Sagittarius A* Event Horizon Telescope Results. I. The Shadow of the Supermassive Black Hole in the Center of the Milky Way}",
	eprint = "2311.08680",
	archivePrefix = "arXiv",
	primaryClass = "astro-ph.HE",
	doi = "10.3847/2041-8213/ac6674",
	journal = "Astrophys. J. Lett.",
	volume = "930",
	number = "2",
	pages = "L12",
	year = "2022"
}

@article{EventHorizonTelescope:2019dse,
	author = "Akiyama, Kazunori and others",
	collaboration = "Event Horizon Telescope",
	title = "{First M87 Event Horizon Telescope Results. I. The Shadow of the Supermassive Black Hole}",
	eprint = "1906.11238",
	archivePrefix = "arXiv",
	primaryClass = "astro-ph.GA",
	doi = "10.3847/2041-8213/ab0ec7",
	journal = "Astrophys. J. Lett.",
	volume = "875",
	pages = "L1",
	year = "2019"
}

@article{PhysRevLett.14.57,
	title = {Gravitational Collapse and Space-Time Singularities},
	author = {Penrose, Roger},
	journal = {Phys. Rev. Lett.},
	volume = {14},
	issue = {3},
	pages = {57--59},
	numpages = {0},
	year = {1965},
	month = {Jan},
	publisher = {American Physical Society},
	doi = {10.1103/PhysRevLett.14.57},
	url = {https://link.aps.org/doi/10.1103/PhysRevLett.14.57}
}

@article{hawking1970singularities,
	title={The singularities of gravitational collapse and cosmology},
	author={Hawking, Stephen William and Penrose, Roger},
	journal={Proceedings of the Royal Society of London. A. Mathematical and Physical Sciences},
	volume={314},
	number={1519},
	pages={529--548},
	year={1970},
	publisher={The Royal Society London}
}

@article{Rov,
	author = {C.\ Rovelli},
	title = "{Quantum Gravity}",
	journal = "Cambridge University Press, Cambridge, UK",
	year = "2004"
}

@article{Thiemann:2001gmi,
	author = "Thiemann, Thomas",
	title = "{Modern canonical quantum general relativity}",
	eprint = "gr-qc/0110034",
	archivePrefix = "arXiv",
	reportNumber = "AEI-2001-119",
	year = "2001"
}

@article{Ashtekar:2004eh,
	author = "Ashtekar, Abhay and Lewandowski, Jerzy",
	title = "{Background independent quantum gravity: A Status report}",
	eprint = "gr-qc/0404018",
	archivePrefix = "arXiv",
	doi = "10.1088/0264-9381/21/15/R01",
	journal = "Class. Quant. Grav.",
	volume = "21",
	pages = "R53",
	year = "2004"
}

@article{Han:2005km,
	author = "Han, Muxin and Huang, Weiming and Ma, Yongge",
	title = "{Fundamental structure of loop quantum gravity}",
	eprint = "gr-qc/0509064",
	archivePrefix = "arXiv",
	doi = "10.1142/S0218271807010894",
	journal = "Int. J. Mod. Phys. D",
	volume = "16",
	pages = "1397--1474",
	year = "2007"
}

@article{Yang:2022aec,
	author = "Yang, Jinsong and Zhang, Cong and Zhang, Xiangdong",
	title = "{Alternative k=-1 loop quantum cosmology}",
	eprint = "2212.05748",
	archivePrefix = "arXiv",
	primaryClass = "gr-qc",
	doi = "10.1103/PhysRevD.107.046012",
	journal = "Phys. Rev. D",
	volume = "107",
	number = "4",
	pages = "046012",
	year = "2023"
}

@article{Han:2024ydv,
	author = "Han, Muxin and Liu, Hongguang and Qu, Dongxue and Vidotto, Francesca and Zhang, Cong",
	title = "{Cosmological dynamics from covariant loop quantum gravity with scalar matter}",
	eprint = "2402.07984",
	archivePrefix = "arXiv",
	primaryClass = "gr-qc",
	doi = "10.1103/PhysRevD.111.086012",
	journal = "Phys. Rev. D",
	volume = "111",
	number = "8",
	pages = "086012",
	year = "2025"
}

@article{Zhang:2024ney,
	author = "Zhang, Cong and Lewandowski, Jerzy and Ma, Yongge and Yang, Jinsong",
	title = "{Black holes and covariance in effective quantum gravity: A solution without Cauchy horizons}",
	eprint = "2412.02487",
	archivePrefix = "arXiv",
	primaryClass = "gr-qc",
	doi = "10.1103/d6ks-d576",
	journal = "Phys. Rev. D",
	volume = "112",
	number = "4",
	pages = "044054",
	year = "2025"
}

@article{Kelly:2020uwj,
	author = "Kelly, Jarod George and Santacruz, Robert and Wilson-Ewing, Edward",
	title = "{Effective loop quantum gravity framework for vacuum spherically symmetric spacetimes}",
	eprint = "2006.09302",
	archivePrefix = "arXiv",
	primaryClass = "gr-qc",
	doi = "10.1103/PhysRevD.102.106024",
	journal = "Phys. Rev. D",
	volume = "102",
	number = "10",
	pages = "106024",
	year = "2020"
}

@article{Lewandowski:2022zce,
	author = "Lewandowski, Jerzy and Ma, Yongge and Yang, Jinsong and Zhang, Cong",
	title = "{Quantum Oppenheimer-Snyder and Swiss Cheese Models}",
	eprint = "2210.02253",
	archivePrefix = "arXiv",
	primaryClass = "gr-qc",
	doi = "10.1103/PhysRevLett.130.101501",
	journal = "Phys. Rev. Lett.",
	volume = "130",
	number = "10",
	pages = "101501",
	year = "2023"
}

@article{Yang:2022btw,
	author = "Yang, Jinsong and Zhang, Cong and Ma, Yongge",
	title = "{Shadow and stability of quantum-corrected black holes}",
	eprint = "2211.04263",
	archivePrefix = "arXiv",
	primaryClass = "gr-qc",
	doi = "10.1140/epjc/s10052-023-11800-8",
	journal = "Eur. Phys. J. C",
	volume = "83",
	number = "7",
	pages = "619",
	year = "2023"
}

@article{Zhang:2023okw,
	author = "Zhang, Cong and Ma, Yongge and Yang, Jinsong",
	title = "{Black hole image encoding quantum gravity information}",
	eprint = "2302.02800",
	archivePrefix = "arXiv",
	primaryClass = "gr-qc",
	month = "2",
	year = "2023"
}

@article{Muniz:2024wiv,
	author = "Muniz, C. R. and Alencar, G. and Cunha, M. S. and Olmo, Gonzalo J.",
	title = "{Static and stationary black bounces inspired by loop quantum gravity}",
	eprint = "2408.08542",
	archivePrefix = "arXiv",
	primaryClass = "gr-qc",
	doi = "10.1103/h7rn-4ht6",
	journal = "Phys. Rev. D",
	volume = "112",
	number = "2",
	pages = "024018",
	year = "2025"
}

@article{Simpson:2018tsi,
	author = "Simpson, Alex and Visser, Matt",
	title = "{Black-bounce to traversable wormhole}",
	eprint = "1812.07114",
	archivePrefix = "arXiv",
	primaryClass = "gr-qc",
	doi = "10.1088/1475-7516/2019/02/042",
	journal = "JCAP",
	volume = "02",
	pages = "042",
	year = "2019"
}

@article{He:2025hbu,
	author = "He, Ke-Jian and Ye, Huan and Zeng, Xiao-Xiong and Li, Li-Fang and Xu, Peng",
	title = "{Shadow and accretion disk images of the rotation loop quantum black bounce*}",
	eprint = "2502.08388",
	archivePrefix = "arXiv",
	primaryClass = "gr-qc",
	doi = "10.1088/1674-1137/adf4a2",
	journal = "Chin. Phys.",
	volume = "49",
	number = "12",
	pages = "125103",
	year = "2025"
}

@article{LIGOScientific:2021sio,
	author = "Abbott, R. and others",
	collaboration = "LIGO Scientific, VIRGO, KAGRA",
	title = "{Tests of General Relativity with GWTC-3}",
	eprint = "2112.06861",
	archivePrefix = "arXiv",
	primaryClass = "gr-qc",
	reportNumber = "LIGO-P2100275",
	month = "12",
	year = "2021"
}

@article{Berti:2015itd,
	author = "Berti, Emanuele and others",
	title = "{Testing General Relativity with Present and Future Astrophysical Observations}",
	eprint = "1501.07274",
	archivePrefix = "arXiv",
	primaryClass = "gr-qc",
	doi = "10.1088/0264-9381/32/24/243001",
	journal = "Class. Quant. Grav.",
	volume = "32",
	pages = "243001",
	year = "2015"
}

@article{Berti:2018vdi,
	author = "Berti, Emanuele and Yagi, Kent and Yang, Huan and Yunes, Nicol\'as",
	title = "{Extreme Gravity Tests with Gravitational Waves from Compact Binary Coalescences: (II) Ringdown}",
	eprint = "1801.03587",
	archivePrefix = "arXiv",
	primaryClass = "gr-qc",
	doi = "10.1007/s10714-018-2372-6",
	journal = "Gen. Rel. Grav.",
	volume = "50",
	number = "5",
	pages = "49",
	year = "2018"
}

@article{Cardoso:2019rvt,
	author = "Cardoso, Vitor and Pani, Paolo",
	title = "{Testing the nature of dark compact objects: a status report}",
	eprint = "1904.05363",
	archivePrefix = "arXiv",
	primaryClass = "gr-qc",
	doi = "10.1007/s41114-019-0020-4",
	journal = "Living Rev. Rel.",
	volume = "22",
	number = "1",
	pages = "4",
	year = "2019"
}

@Article{Fu:2022cul,
	author        = {Fu, Guoyang and Zhang, Dan and Liu, Peng and Kuang, Xiao-Mei and Pan, Qiyuan and Wu, Jian-Pin},
	journal       = {Phys. Rev. D},
	title         = {{Quasinormal modes and Hawking radiation of a charged Weyl black hole}},
	year          = {2023},
	number        = {4},
	pages         = {044049},
	volume        = {107},
	archiveprefix = {arXiv},
	doi           = {10.1103/PhysRevD.107.044049},
	eprint        = {2207.12927},
	primaryclass  = {gr-qc},
}

@Article{Fu:2023drp,
	author        = {Fu, Guoyang and Zhang, Dan and Liu, Peng and Kuang, Xiao-Mei and Wu, Jian-Pin},
	journal       = {Phys. Rev. D},
	title         = {{Peculiar properties in quasinormal spectra from loop quantum gravity effect}},
	year          = {2024},
	number        = {2},
	pages         = {026010},
	volume        = {109},
	archiveprefix = {arXiv},
	doi           = {10.1103/PhysRevD.109.026010},
	eprint        = {2301.08421},
	primaryclass  = {gr-qc},
}

@Article{Gong:2023ghh,
	author        = {Gong, Huajie and Li, Shulan and Zhang, Dan and Fu, Guoyang and Wu, Jian-Pin},
	title         = {{Quasinormal modes of quantum-corrected black holes}},
	year          = {2023},
	month         = {12},
	archiveprefix = {arXiv},
	eprint        = {2312.17639},
	primaryclass  = {gr-qc},
}

@article{Yang:2024lmj,
	author = "Yang, Sen and Zhang, Yu-Peng and Zhu, Tao and Zhao, Li and Liu, Yu-Xiao",
	title = "{Gravitational waveforms from periodic orbits around a quantum-corrected black hole}",
	eprint = "2407.00283",
	archivePrefix = "arXiv",
	primaryClass = "gr-qc",
	doi = "10.1088/1475-7516/2025/01/091",
	journal = "JCAP",
	volume = "01",
	pages = "091",
	year = "2025"
}

@article{Zhang:2024nny,
	author = "Zhang, Dan and Gong, Huajie and Fu, Guoyang and Wu, Jian-Pin and Pan, Qiyuan",
	title = "{Quasinormal modes of a regular black hole with sub-Planckian curvature}",
	eprint = "2402.15085",
	archivePrefix = "arXiv",
	primaryClass = "gr-qc",
	doi = "10.1140/epjc/s10052-024-12928-x",
	journal = "Eur. Phys. J. C",
	volume = "84",
	number = "6",
	pages = "564",
	year = "2024"
}

@article{Song:2024kkx,
	author = "Song, Zhijun and Gong, Huajie and Li, Hai-Li and Fu, Guoyang and Zhu, Li-Gang and Wu, Jian-Pin",
	title = "{Quasinormal modes and ringdown waveform of the Frolov black hole}",
	eprint = "2406.04787",
	archivePrefix = "arXiv",
	primaryClass = "gr-qc",
	month = "6",
	year = "2024"
}

@article{Xia:2023zlf,
	author = "Xia, Zhong-Wu and Yang, Hao and Miao, Yan-Gang",
	title = "{Scalar fields around a rotating loop quantum gravity black hole: waveform, quasi-normal modes and superradiance}",
	eprint = "2310.00253",
	archivePrefix = "arXiv",
	primaryClass = "gr-qc",
	doi = "10.1088/1361-6382/ad6129",
	journal = "Class. Quant. Grav.",
	volume = "41",
	number = "16",
	pages = "165010",
	year = "2024"
}

@article{Guo:2023nkd,
	author = "Guo, Wen-Di and Tan, Qin and Liu, Yu-Xiao",
	title = "{Quasinormal modes and greybody factor of a Lorentz-violating black hole}",
	eprint = "2312.16605",
	archivePrefix = "arXiv",
	primaryClass = "gr-qc",
	doi = "10.1088/1475-7516/2024/07/008",
	journal = "JCAP",
	volume = "07",
	pages = "008",
	year = "2024"
}

@article{Yang:2024prm,
	author = "Yang, Hao and Xia, Zhong-Wu and Miao, Yan-Gang",
	title = "{Echoes and quasi-normal modes of perturbations around Schwarzschild traversable wormholes}",
	eprint = "2406.00377",
	archivePrefix = "arXiv",
	primaryClass = "gr-qc",
	doi = "10.1140/epjc/s10052-025-14446-w",
	journal = "Eur. Phys. J. C",
	volume = "85",
	number = "7",
	pages = "742",
	year = "2025"
}

@article{Tan:2024qij,
	author = "Tan, Qin and Long, Sheng and Deng, Weike and Jing, Jiliang",
	title = "{Quasinormal modes and echoes of a double braneworld}",
	eprint = "2410.06945",
	archivePrefix = "arXiv",
	primaryClass = "gr-qc",
	doi = "10.1007/JHEP02(2025)055",
	journal = "JHEP",
	volume = "02",
	pages = "055",
	year = "2025"
}

@article{Dong:2024ams,
	author = "Dong, Zhongzhinan and Zhang, Dan and Fu, Guoyang and Wu, Jian-Pin",
	title = "{Quasinormal modes of a d-dimensional regular black hole featuring an integrable singularity}",
	eprint = "2412.20457",
	archivePrefix = "arXiv",
	primaryClass = "gr-qc",
	doi = "10.1140/epjc/s10052-025-13926-3",
	journal = "Eur. Phys. J. C",
	volume = "85",
	number = "2",
	pages = "215",
	year = "2025"
}

@article{Deng:2025hfn,
    author = "Deng, Weike and Long, Sheng and Tan, Qin and Chen, Zu-Cheng and Jing, Jiliang",
    title = "{Scalar-gravitational quasinormal modes and echoes in a five dimensional thick brane}",
    eprint = "2508.20937",
    archivePrefix = "arXiv",
    primaryClass = "gr-qc",
    doi = "10.1007/JHEP01(2026)066",
    journal = "JHEP",
    volume = "01",
    pages = "066",
    year = "2026"
}

@article{Cardoso:2016oxy,
	author = "Cardoso, Vitor and Hopper, Seth and Macedo, Caio F. B. and Palenzuela, Carlos and Pani, Paolo",
	title = "{Gravitational-wave signatures of exotic compact objects and of quantum corrections at the horizon scale}",
	eprint = "1608.08637",
	archivePrefix = "arXiv",
	primaryClass = "gr-qc",
	doi = "10.1103/PhysRevD.94.084031",
	journal = "Phys. Rev. D",
	volume = "94",
	number = "8",
	pages = "084031",
	year = "2016"
}

@article{Hui:2023ibl,
	author = "Hui, Siyuan and Mu, Benrong and Wang, Peng",
	title = "{Echoes from charged black holes influenced by quintessence}",
	eprint = "2305.11200",
	archivePrefix = "arXiv",
	primaryClass = "gr-qc",
	doi = "10.1016/j.dark.2023.101396",
	journal = "Phys. Dark Univ.",
	volume = "43",
	pages = "101396",
	year = "2024"
}

@article{Zhang:2023mzb,
	author = "Zhang, Chen and Gao, Yong and Xia, Cheng-Jun and Xu, Renxin",
	title = "{Rescaling strange-cluster stars and its implications on gravitational-wave echoes}",
	eprint = "2305.13323",
	archivePrefix = "arXiv",
	primaryClass = "astro-ph.HE",
	doi = "10.1103/PhysRevD.108.063002",
	journal = "Phys. Rev. D",
	volume = "108",
	number = "6",
	pages = "063002",
	year = "2023"
}

@article{Yang:2022ryf,
	author = "Yang, Yi and Liu, Dong and Xu, Zhaoyi and Long, Zheng-Wen",
	title = "{Ringing and echoes from black bounces surrounded by the string cloud}",
	eprint = "2210.12641",
	archivePrefix = "arXiv",
	primaryClass = "gr-qc",
	doi = "10.1140/epjc/s10052-023-11382-5",
	journal = "Eur. Phys. J. C",
	volume = "83",
	number = "3",
	pages = "217",
	year = "2023"
}

@article{Liu:2021aqh,
	author = "Liu, Hang and Qian, Wei-Liang and Liu, Yunqi and Wu, Jian-Pin and Wang, Bin and Yue, Rui-Hong",
	title = "{Alternative mechanism for black hole echoes}",
	eprint = "2104.11912",
	archivePrefix = "arXiv",
	primaryClass = "gr-qc",
	doi = "10.1103/PhysRevD.104.044012",
	journal = "Phys. Rev. D",
	volume = "104",
	number = "4",
	pages = "044012",
	year = "2021"
}

@article{Abedi:2016hgu,
	author = "Abedi, Jahed and Dykaar, Hannah and Afshordi, Niayesh",
	title = "{Echoes from the Abyss: Tentative evidence for Planck-scale structure at black hole horizons}",
	eprint = "1612.00266",
	archivePrefix = "arXiv",
	primaryClass = "gr-qc",
	doi = "10.1103/PhysRevD.96.082004",
	journal = "Phys. Rev. D",
	volume = "96",
	number = "8",
	pages = "082004",
	year = "2017"
}

@article{Conklin:2017lwb,
	author = "Conklin, Randy S. and Holdom, Bob and Ren, Jing",
	title = "{Gravitational wave echoes through new windows}",
	eprint = "1712.06517",
	archivePrefix = "arXiv",
	primaryClass = "gr-qc",
	doi = "10.1103/PhysRevD.98.044021",
	journal = "Phys. Rev. D",
	volume = "98",
	number = "4",
	pages = "044021",
	year = "2018"
}

@article{Cardoso:2017cqb,
	author = "Cardoso, Vitor and Pani, Paolo",
	title = "{Tests for the existence of black holes through gravitational wave echoes}",
	eprint = "1709.01525",
	archivePrefix = "arXiv",
	primaryClass = "gr-qc",
	doi = "10.1038/s41550-017-0225-y",
	journal = "Nature Astron.",
	volume = "1",
	number = "9",
	pages = "586--591",
	year = "2017"
}

@article{Mark:2017dnq,
	author = "Mark, Zachary and Zimmerman, Aaron and Du, Song Ming and Chen, Yanbei",
	title = "{A recipe for echoes from exotic compact objects}",
	eprint = "1706.06155",
	archivePrefix = "arXiv",
	primaryClass = "gr-qc",
	reportNumber = "LIGO-P1700145",
	doi = "10.1103/PhysRevD.96.084002",
	journal = "Phys. Rev. D",
	volume = "96",
	number = "8",
	pages = "084002",
	year = "2017"
}

@article{Konoplya:2018yrp,
	author = "Konoplya, R. A. and Stuchl\'\i{}k, Z. and Zhidenko, A.",
	title = "{Echoes of compact objects: new physics near the surface and matter at a distance}",
	eprint = "1810.01295",
	archivePrefix = "arXiv",
	primaryClass = "gr-qc",
	doi = "10.1103/PhysRevD.99.024007",
	journal = "Phys. Rev. D",
	volume = "99",
	number = "2",
	pages = "024007",
	year = "2019"
}

@article{LongoMicchi:2019wsh,
	author = "Longo Micchi, Luis Felipe and Chirenti, Cecilia",
	title = "{Spicing up the recipe for echoes from exotic compact objects: orbital differences and corrections in rotating backgrounds}",
	eprint = "1912.05419",
	archivePrefix = "arXiv",
	primaryClass = "gr-qc",
	doi = "10.1103/PhysRevD.101.084010",
	journal = "Phys. Rev. D",
	volume = "101",
	number = "8",
	pages = "084010",
	year = "2020"
}

@article{Churilova:2019cyt,
	author = "Churilova, M. S. and Stuchlik, Z.",
	title = "{Ringing of the regular black-hole/wormhole transition}",
	eprint = "1911.11823",
	archivePrefix = "arXiv",
	primaryClass = "gr-qc",
	doi = "10.1088/1361-6382/ab7717",
	journal = "Class. Quant. Grav.",
	volume = "37",
	number = "7",
	pages = "075014",
	year = "2020"
}

@article{Bronnikov:2019sbx,
	author = "Bronnikov, Kirill A. and Konoplya, Roman A.",
	title = "{Echoes in brane worlds: ringing at a black hole--wormhole transition}",
	eprint = "1912.05315",
	archivePrefix = "arXiv",
	primaryClass = "gr-qc",
	doi = "10.1103/PhysRevD.101.064004",
	journal = "Phys. Rev. D",
	volume = "101",
	number = "6",
	pages = "064004",
	year = "2020"
}

@article{DuttaRoy:2019hij,
	author = "Dutta Roy, Poulami and Aneesh, S. and Kar, Sayan",
	title = "{Revisiting a family of wormholes: geometry, matter, scalar quasinormal modes and echoes}",
	eprint = "1910.08746",
	archivePrefix = "arXiv",
	primaryClass = "gr-qc",
	doi = "10.1140/epjc/s10052-020-8409-5",
	journal = "Eur. Phys. J. C",
	volume = "80",
	number = "9",
	pages = "850",
	year = "2020"
}

@article{Chowdhury:2020rfj,
	author = "Chowdhury, Avijit and Banerjee, Narayan",
	title = "{Echoes from a singularity}",
	eprint = "2006.16522",
	archivePrefix = "arXiv",
	primaryClass = "gr-qc",
	doi = "10.1103/PhysRevD.102.124051",
	journal = "Phys. Rev. D",
	volume = "102",
	number = "12",
	pages = "124051",
	year = "2020"
}

@article{Chowdhury:2022zqg,
	author = "Chowdhury, Avijit and Devi, Saraswati and Chakrabarti, Sayan",
	title = "{Naked singularity in 4D Einstein-Gauss-Bonnet novel gravity: Echoes and instability}",
	eprint = "2202.13698",
	archivePrefix = "arXiv",
	primaryClass = "gr-qc",
	doi = "10.1103/PhysRevD.106.024023",
	journal = "Phys. Rev. D",
	volume = "106",
	number = "2",
	pages = "024023",
	year = "2022"
}

@article{Zhang:2025ygb,
	author = "Zhang, Dan and Tan, Qin and Fu, Guoyang and Gong, Huajie and Wu, Jian-Pin and Pan, Qiyuan",
	title = "{Echoes from the Minkowski-core spacetime}",
	eprint = "2509.23215",
	archivePrefix = "arXiv",
	primaryClass = "gr-qc",
	doi = "10.1007/s11433-025-2904-2",
	journal = "Sci. China Phys. Mech. Astron.",
	volume = "69",
	number = "5",
	pages = "250412",
	year = "2026"
}

@article{Abdalla:2010nq,
	author = "Abdalla, E. and Pellicer, C. E. and de Oliveira, Jeferson and Pavan, A. B.",
	title = {{Phase transitions and regions of stability in Reissner-Nordstr\"om holographic superconductors}},
	eprint = "1010.2806",
	archivePrefix = "arXiv",
	primaryClass = "hep-th",
	doi = "10.1103/PhysRevD.82.124033",
	journal = "Phys. Rev. D",
	volume = "82",
	pages = "124033",
	year = "2010"
}

@article{Zhu:2014sya,
	author = "Zhu, Zhiying and Zhang, Shao-Jun and Pellicer, C. E. and Wang, Bin and Abdalla, Elcio",
	title = {{Stability of Reissner-Nordstr\"om black hole in de Sitter background under charged scalar perturbation}},
	eprint = "1405.4931",
	archivePrefix = "arXiv",
	primaryClass = "hep-th",
	doi = "10.1103/PhysRevD.90.044042",
	journal = "Phys. Rev. D",
	volume = "90",
	number = "4",
	pages = "044042",
	year = "2014",
	note = "[Addendum: Phys. Rev. D 90 (2014) 049904 ]"
}

@article{Lin:2022rtx,
	author = "Lin, Kai and Qian, Wei-Liang",
	title = "{Echoes of axial gravitational perturbations in stars of uniform density}",
	eprint = "2204.09531",
	archivePrefix = "arXiv",
	primaryClass = "gr-qc",
	doi = "10.1088/1674-1137/acd681",
	journal = "Chin. Phys. C",
	volume = "47",
	number = "8",
	pages = "085101",
	year = "2023"
}

@article{Berti:2007dg,
	author = "Berti, Emanuele and Cardoso, Vitor and Gonzalez, Jose A. and Sperhake, Ulrich",
	title = "{Mining information from binary black hole mergers: A Comparison of estimation methods for complex exponentials in noise}",
	eprint = "gr-qc/0701086",
	archivePrefix = "arXiv",
	doi = "10.1103/PhysRevD.75.124017",
	journal = "Phys. Rev. D",
	volume = "75",
	pages = "124017",
	year = "2007"
}

@article{Konoplya:2011qq,
	author = "Konoplya, R. A. and Zhidenko, A.",
	title = "{Quasinormal modes of black holes: From astrophysics to string theory}",
	eprint = "1102.4014",
	archivePrefix = "arXiv",
	primaryClass = "gr-qc",
	doi = "10.1103/RevModPhys.83.793",
	journal = "Rev. Mod. Phys.",
	volume = "83",
	pages = "793--836",
	year = "2011"
}

@article{Ferrari:2007rc,
	author = "Ferrari, Valeria. and Gualtieri, Leonardo and Marassi, Stefania",
	title = "{A New approach to the study of quasi-normal modes of rotating stars}",
	eprint = "0709.2925",
	archivePrefix = "arXiv",
	primaryClass = "gr-qc",
	doi = "10.1103/PhysRevD.76.104033",
	journal = "Phys. Rev. D",
	volume = "76",
	pages = "104033",
	year = "2007"
}

@article{Pani:2012bp,
	author = "Pani, Paolo and Cardoso, Vitor and Gualtieri, Leonardo and Berti, Emanuele and Ishibashi, Akihiro",
	title = "{Perturbations of slowly rotating black holes: massive vector fields in the Kerr metric}",
	eprint = "1209.0773",
	archivePrefix = "arXiv",
	primaryClass = "gr-qc",
	doi = "10.1103/PhysRevD.86.104017",
	journal = "Phys. Rev. D",
	volume = "86",
	pages = "104017",
	year = "2012"
}

@article{Pierini:2023btw,
	author = "Pierini, Lorenzo",
	title = "{Quasinormal modes of black holes in Einstein-dilaton Gauss-Bonnet gravity}",
	journal = "Rome U.",
	year = "2023"
}

@article{Wu:2022eiv,
	author = "Wu, S. R. and Wang, B. Q. and Liu, Dong and Long, Z. W.",
	title = "{Echoes of charged black-bounce spacetimes}",
	eprint = "2201.08415",
	archivePrefix = "arXiv",
	primaryClass = "gr-qc",
	doi = "10.1140/epjc/s10052-022-10938-1",
	journal = "Eur. Phys. J. C",
	volume = "82",
	number = "11",
	pages = "998",
	year = "2022"
}

@article{Yang:2021cvh,
	author = "Yang, Yi and Liu, Dong and Xu, Zhaoyi and Xing, Yujia and Wu, Shurui and Long, Zheng-Wen",
	title = "{Echoes of novel black-bounce spacetimes}",
	eprint = "2107.06554",
	archivePrefix = "arXiv",
	primaryClass = "gr-qc",
	doi = "10.1103/PhysRevD.104.104021",
	journal = "Phys. Rev. D",
	volume = "104",
	number = "10",
	pages = "104021",
	year = "2021"
}

@article{Moderski:2005hf,
	author = "Moderski, Rafal and Rogatko, Marek",
	title = "{Evolution of a self-interacting scalar field in the spacetime of a higher dimensional black hole}",
	eprint = "hep-th/0508175",
	archivePrefix = "arXiv",
	doi = "10.1103/PhysRevD.72.044027",
	journal = "Phys. Rev. D",
	volume = "72",
	pages = "044027",
	year = "2005"
}

@article{Moderski:2001gt,
	author = "Moderski, Rafa and Rogatko, Marek",
	title = "{Late time evolution of a charged massless scalar field in the space-time of a dilaton black hole}",
	doi = "10.1103/PhysRevD.63.084014",
	journal = "Phys. Rev. D",
	volume = "63",
	pages = "084014",
	year = "2001"
}

@article{Moderski:2001tk,
	author = "Moderski, Rafal and Rogatko, Marek",
	title = "{Late time evolution of a selfinteracting scalar field in the space-time of dilaton black hole}",
	eprint = "gr-qc/0105056",
	archivePrefix = "arXiv",
	doi = "10.1103/PhysRevD.64.044024",
	journal = "Phys. Rev. D",
	volume = "64",
	pages = "044024",
	year = "2001"
}
\end{document}